\documentclass[]{jfm}          
\usepackage[latin1]{inputenc}
\usepackage{graphicx}
\usepackage{amsmath}
\usepackage{amssymb}
\usepackage{color}
\usepackage[dvips]{epsfig}
\usepackage{epstopdf}
\usepackage{times,mathptmx}
\usepackage{epstopdf}
\usepackage{natbib}
\usepackage{color}
\renewcommand{\Re}{\mathrm{Re}}

\begin{document}


\shortauthor{A. L. Dubov et al} 
\title{Inertial focusing of finite-size particles in microchannels}

\author[E. S. Asmolov et al]
{Evgeny S. Asmolov$^{1,2}$\thanks{Email address for correspondence: aes50@yandex.ru}, Alexander L. Dubov$^{1}$, Tatiana V. Nizkaya$^{1}$,  Jens Harting$^{2,3,4}$, and Olga I. Vinogradova$^{1,5,6}$\thanks{Email address for correspondence: oivinograd@yahoo.com}
}

\affiliation{
$^1$A.N.~Frumkin Institute of Physical Chemistry and Electrochemistry,\\[\affilskip]
Russian Academy of Sciences, 31 Leninsky Prospect, 119071 Moscow, Russia\\[\affilskip]
$^2$Institute of Mechanics, M. V. Lomonosov Moscow State University, 119991 Moscow, Russia\\[\affilskip]
$^3$Helmholtz Institute Erlangen-N\"urnberg for Renewable Energy, Forschungszentrum J\"ulich,\\ F\"urther Str. 248, 90429 N\"{u}rnberg, Germany\\[\affilskip]
$^4$Department of Applied Physics, Eindhoven University of Technology, PO box 513, 5600MB Eindhoven, The Netherlands\\[\affilskip]
$^5$ Faculty of Science and Technology, University of Twente, 7500 AE Enschede, The Netherlands\\[\affilskip]
$^6$Department of Physics, M. V. Lomonosov Moscow State University, 119991
Moscow, Russia\\[\affilskip]
$^7$DWI - Leibniz Institute for Interactive Materials, Forckenbeckstr. 50, 52056 Aachen, Germany\\[\affilskip]
}
\date{Received: date / Accepted: date}

\maketitle

\begin{abstract}

At finite Reynolds numbers, $\Re$, particles migrate across laminar flow streamlines  to their equilibrium positions in microchannels. This
migration is attributed to a lift force, and the balance between this lift and gravity determines the location of particles in channels. Here we demonstrate that velocity of finite-size particles located near a channel wall differs significantly from that of an undisturbed flow, and that their equilibrium position depends on this, referred to as slip velocity, difference. We then present theoretical arguments, which allow us to generalize expressions for a lift force, originally suggested for some limiting cases and $\Re \ll 1$, to finite-size particles in a channel flow at $\Re \leq 20$.
Our theoretical model, validated by lattice Boltzmann simulations, provides considerable insight into inertial migration of finite-size particles in microchannel and suggests some novel microfluidic approaches to separate them by size or density at a moderate $\Re$.

\end{abstract}

\section{Introduction}
\label{sec_intro}

Microfluidic systems have been shown to be very useful
for continuous manipulation and separation of microparticles with increased control and sensitivity, which is important for a wide range
of applications in chemistry, biology, and medicine. Traditional microfluidic techniques of particle manipulation rely on low Reynolds number laminar flow. Under these conditions, when no external forces are applied, particles follow fluid streamlines. Contrary to this, particles migrate  across streamlines to some stationary positions in microchannels when inertial aspects of the flow become significant. This migration is attributed to inertial lift forces, which are currently successfully used in microfluidic systems to focus and separate particles of different
sizes continuously, at high flow rate, and without external forces~\citep{dicarlo07,bhagat08}.
The rapid development of inertial microfluidics has raised a considerable interest in the
lift forces on particles in confined flows. We mention below what we believe are the most relevant contributions.

Inertial lift forces on neutrally buoyant particles have been originally reported for
macroscopic channels~\citep{Segre:Silb62b}. This pioneering work has concluded that particles
focus to a narrow annulus at radial position $0.6$ of a pipe radius, and argued that lift forces vanish at this equilibrium position. However, no particle manipulation systems have been
explored based on macroscale systems. Much later this inertial focusing has provided the basis for various methods of particle separation by size
or shape in microfluidics devices (see \citet{martel2014inertial} and \citet{zhang2016} for recent
reviews). In these microfluidic applications the inertial lift has been balanced by the Dean force due to a secondary rotational
flow caused by inertia of the fluid itself, which can be generated in curved channels~\citep{bhagat08}. These Dean drag forces alter
equilibrium positions of particles. The preferred location of particles in microchannels could also be controlled by the
balance between inertial lift and external forces, such as electric~\citep{zhang2014real} or
magnetic~\citep{dutz2017fractionation}.

In recent years extensive efforts have gone into experimental investigating particle equilibrium positions in cylindrical~\citep{Matas:etal:JFM04,morita2017} and rectangular channels~\citep{choi2011,miura2014,hood2016}.  \citet{Matas:etal:JFM04} have shown that the Segr\'{e}-Silberberg annulus for neutrally-buoyant particles shifts toward the wall as $\Re$ increases and toward the pipe center as particle size increases. At large $\Re\geq 600$, some particles accumulate in an inner
annulus near the pipe centre. \citet{morita2017} have found that the inner annulus is not a true equilibrium position, but a transient zone, and that in a long enough pipe  all particles accumulate within the Segr\'{e}-Silberberg annulus. It has also been found that equilibrium positions of slightly non neutrally-buoyant particles in a horizontal pipe are shifted toward a pipe bottom~\citep{Matas:etal:JFM04}.

During last several years numerical calculations~\citep{dicarlo2009prl,liu2015,Loisel2015} and computer simulations~\citep{Chun:Ladd06,kilimnik2011} have also been concerned with phenomena of the inertial migration. It has been shown that in rectangular channels particles initially migrate rapidly to manifolds, and then slowly focus within the manifolds to stable equilibrium positions near wall centers and  channel corners~\citep{Chun:Ladd06,dicarlo2009prl,hood2016}. There could be two, four or eight equilibrium positions depending on the particle size, channel aspect ratio and Reynolds number. Overall, simulations are consistent with experimental results~\citep{choi2011,miura2014,hood2016}.

 There is also a large literature describing attempts to provide a theory of inertial lift.  An asymptotic approach, which can shed light on these phenomena, has been developed by several authors~\citep{Saffman65,Ho:Leal74,Vass:Cox76,Cox:Hsu77,Schon:Hinch89,
Asmolov99,Matas:etal:JFM04,matas2009}. Most papers have considered a plane Poiseuille flow except the work by~\citet{matas2009} where a pipe flow has been addressed. The approach can be applied when the particle Reynolds number, $\mathrm{Re}_{p}=a^{2}G/\nu$,  where $a$ is
the particle radius, $G$ is the characteristic shear rate and $\nu$ is the
kinematic viscosity, is small. If so, to the leading order in $\mathrm{Re}_{p}$, the disturbance
flow is governed by the Stokes equations, and a spherical particle experiences
a drag and a torque, but no lift.  The Stokeslet disturbance originates from the particle
translational motion relative to the fluid and is proportional to the slip velocity $V'_s=V'-U'$, where $V'$ and $U'$ are forward velocities of the particle and of the undisturbed flow
at the particle center. The stresslet is induced by free rotation of the sphere in the shear flow and is proportional to $G$.
The lift force has then been deduced from the solution of the
next-order equations which accounts a non-linear coupling between the two disturbances~\citep{Vass:Cox76}:

\begin{equation}
F_{l}^{\prime }=\rho a^{2}\left( c_{l0}{a^{2}}G^{2}+c_{l1}aGV_{s}^{\prime
}+c_{l2}V_{s}^{\prime 2}\right) ,
\label{cher}
\end{equation}%
where $\rho $ is the fluid density.
The coefficients $c_{li}$ ($i=0,1,2$) generally depend
on several dimensionless parameters, such as $z/a$, $H/a$, $V_s'/U_{m}'$, and on the channel Reynolds
number, $\mathrm{Re}=U_{m}'H/\nu$, where $z$ is the distance to the closest
wall, $H$ is the channel thickness, and $U'_m$ is the maximum velocity of the
channel flow. Solutions for $c_l$ have been obtained in some limiting cases only,
and no general analytical equations have still been proposed for finite-sized particles in the channel. Thus, \citet{Vass:Cox76} have calculated the coefficients $c_{l0}^{VC},c_{l1}^{VC},c_{l2}^{VC}$ for pointlike particles
at small channel Reynolds numbers, $\Re \ll 1$, which depend on $z/H$ only and are applicable when $z\gg a$.
\label{add2} \citet{Cheruk:Mclau94} have later evaluated the coefficients $c_{li}^{CM}(z/a)$ for finite-size particles near a single wall in a linear shear flow assuming that $z\sim a$ and proposed simple fits for them. However, it remains unclear if and how earlier theoretical results for pointlike particles at $\Re \ll 1$ or for finite-size particles near a single wall can be generalized to predict the lift of finite-size particles at any $z$ and a finite $\Re$ of a microfluidic channel.

According to Equation~(\ref{cher}) the contribution of the slip velocity to the lift forces dominates when $V_s'\gg Ga$. Since the slip velocity is induced by external forces, such as gravity, it is believed that it impacts a hydrodynamic lift only in the case of non-neutrally buoyant particles. For neutrally buoyant particles with equal to $\rho$ density, the slip velocity is normally considered to be negligibly small~\citep{Ho:Leal74,hood2015}. A corollary from that would be that the lift of neutrally buoyant particles could be due to the stresslet only. Such a conclusion, however, can be justified theoretically only for small particles far from walls, $z \gg a$, but hydrodynamic interactions at finite distances $z\sim a$ can induce a finite slip, $V_s'\sim Ga$, so that all terms in Equation~(\ref{cher}) become comparable~\citep{Cheruk:Mclau94}. The variation of the slip velocity of neutrally buoyant particles in a thin near-wall layer can impact the lift force, but we are unaware of any previous work that has addressed this question.

The purpose of this introduction has been to show that, in spite of its importance for inertial microfluidics, the lift forces of finite-size particles in a bounded geometry of a microchannel still remain poorly understood. In particular, there is still a lack of simple analytical formulas quantifying the lift, as well as of general solutions valid in the large range of parameters typical for real microfluidic devices.
 Given the current upsurge of interest in the inertial hydrodynamic phenomena and their applications to separation of particles in microfluidic devices it would seem timely to provide a more satisfactory theory of a hydrodynamic lift in a microchannel and also to bring some of modern simulation techniques to bear on this problem. In this paper we present some results of a study of a migration of finite-size particles at moderate
channel Reynolds numbers, $\mathrm{Re}\sim 10$, with the special focus on the role of the slip
velocity in the hydrodynamic lift.

Our paper is arranged as follows. In \S 2 we propose a general expression for the lift force on a neutrally buoyant
particle in a microchannel, which reduces to earlier theoretical results~\citep{Vass:Cox76,Cheruk:Mclau94} in relevant limiting cases. We also extend our expression to the case of slightly non-neutrally buoyant particles with the slip velocity smaller than $G_ma$. To access the validity of the proposed theory we use a simulation method described in \S 3, and the numerical results are presented in \S 4.
We conclude in \S 5 with a discussion of our results and their possible relevance for a fractionation of particles in microfluidic devices. Appendices~\ref{slip} and~\ref{app_b} contain a summary of early calculations of lift coefficients and the derivation of differential equations that determine trajectories of particles.

\begin{figure}
  \centering
  \includegraphics[height=5.6cm]{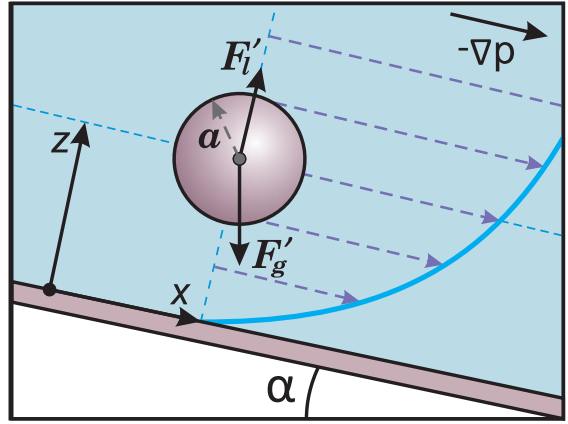}\\
  \caption{Sketch of a migration of a particle of radius $a$ to an equilibrium position in a pressure-driven flow. The locus of this position is determined by the balance between lift, $F_l'$, and gravity, $F_g'$, forces.}
  \label{fig_sketch}
\end{figure}

\section{Theory}
\label{sec_theo}

In this section we propose an analytical expression for the lift force on neutrally buoyant and slightly non-neutrally buoyant particles of radius $a$, which translate parallel to a channel wall. Our expression is valid for $a/H \ll 1$ at any distance $z$ from the channel wall.

We consider a pressure-driven flow in a flat inclined microchannel of thickness $H$. An inclination angle $\alpha\geq0$
is defined relative to the horizontal. The coordinate axis $x$ is
parallel to the channel wall, and the normal to the wall coordinate is denoted by $z$. The geometry is shown in
Figure~\ref{fig_sketch}. The undisturbed velocity profile in such a channel is given by

\begin{equation}
U'(z)=4U_m'z\left( 1-z/H\right) /H.
\label{eq:Uflow}
\end{equation}

Let us now introduce a
dimensionless slip velocity $V_s=V_s'/(a G_{m})$, where $G_m=4U'_m/H$ is the maximum shear rate at the
channel wall. We can then rewrite Equation~(\ref{cher}) as
\begin{equation}
F_{l}'={\rho a^{4}} G_m^{2}c_{l},
  \label{f_scale}
\end{equation}%
with the lift coefficient
\begin{equation}
c_{l}=c_{l0}+c_{l1}V_{s}+c_{l2}V_{s}^{2},
\label{eq_force1}
\end{equation}
which depends on the slip velocity, $V_s$, which in turns can be determined from the Stokes equations (a zero-order solution). Therefore, to construct a general expression for a lift force acting on finite size particles in the channel it is necessary to estimate $V_s$ as a function of $z$.


We begin by studying the classical case of neutrally buoyant (i.e. force- and a torque-free) particles with a density $\rho_p$ equal to that of liquid, $\rho$. The expression for $V_s$  in a linear shear flow near a single wall has been derived before~\citep{Goldman1967} and can be used to calculate the slip velocity in the near-wall region of our channel. The fits for $V_s$ are given in Appendix~\ref{slip}, Equations~(\ref{V_sh})-(\ref{wak}). We first note that depending on $z/a$ one can distinguish between two different regimes of behavior of $V_s$. In the central part of the channel, i.e. when $z/a \gg 1$, the slip contribution to the lift decays as $(a/z)^{3}$~\citep{wakiya1967}, being always very small, but finite. In contrast, when the gap between the sphere and the wall is small, $z/a-1\ll 1$, the slip velocity varies very rapidly with $z/a$~\citep{Goldman1967}:
\begin{equation}
V_s^{nb}=-1+\frac{0.7431}{0.6376-0.200\log \left( z/a-1\right)}.
\label{log}
\end{equation}
As a side note we should like to mention here that a logarithmic singularity in Equation~(\ref{log}) implies that in the near-wall region the lift coefficient, Equation~(\ref{eq_force1}), cannot be fitted by any power law $(a/z)^n$ as it has been previously suggested~\citep{dicarlo2009prl,hood2015,liu2016}.

It follows from Equation~(\ref{log}) that for an immobile particle in a contact with the wall, $z=a$, the slip velocity is largest, $V_{s}=-1$. In this limiting case the lift coefficient also takes its maximum value, $c_{l}^{KL}\simeq 9.257$~\citep{krishnan1995}. Far from the wall, the slip velocity is much smaller and can be neglected, so that we can consider $c_l \simeq c_{l0}$. Therefore, when $a\ll z\ll H$, the value of $c_l$ in Equation~(\ref{eq_force1}) is equal to $c_{l0}^{CV}|_{z/H\rightarrow 0}=55\pi /96\simeq 1.8$~\citep{Cox:Hsu77}, i.e. it becomes much smaller than for a particle at the wall. This illustrates that $c_l$ varies significantly in the vicinity of the wall due to a finite slip.

We now remark that the Stokeslet contribution (the second and the third terms in Equation~(\ref{eq_force1})) is
finite for $z\sim a$ only and vanishes in the central part of the channel. Within the close proximity to the wall we may neglect the corrections to the slip and the lift of order $a/H$ due to parabolic flow~\citep{Pasol2006,yahiaoui2010}
and due to the second wall. Therefore, in this region one can use the results by \citet{Cheruk:Mclau94} for the lift coefficients $c_{li}^{CM}$.
The stresslet contribution to the lift (first term in Equation~(\ref{eq_force1})) is finite
for any $z$. Close to the wall, the effect of particle size for this term is negligible as the coefficient
$c_{l0}^{CM}(z/a)$ is nearly constant~\citep{Cheruk:Mclau94}. So we may describe the stresslet contribution by the coefficient $c_{l0}^{VC}$  obtained by \citet{Vass:Cox76}.
This enables us to construct the following formula for the lift coefficient:
\begin{equation}
c_{l}=c_{l0}^{VC}(z/H)+\gamma c_{l1}^{CM}(z/a)V_{s}+c_{l2}^{CM}(z/a)V_{s}^{2},  \label{our_fit}
\end{equation}%
where $\gamma =G(z)/G_m= 1-2z/H\leq 1$ is a dimensionless local shear rate at
the particle position. The fitting expressions for three lift coefficients are summarized in Appendix~\ref{slip}. We, therefore, use Equation~(\ref{cl0}) to calculate $c_{l0}^{VC}$, Equation~(\ref{cl1CM}) to calculate $c_{l1}^{CM}$, and Equation~(\ref{cl2CM}) for $c_{l2}^{CM}$. Note that in the second term of Equation~(\ref{our_fit}) we have introduced a correction factor $\gamma$, which takes into account the variation
of $G$ in the second term of Equation~(\ref{cher}) and ensures the lift to remain
zero at the channel centerline.

We recall, that Equation~(\ref{our_fit}) is asymptotically
valid for any $z$ when $a/H\ll 1$ and $\Re\ll 1$. However, one can argue that it should be accurate enough at moderate Reynolds numbers. \label{add6} Indeed, the contribution of undisturbed flow to inertial terms in the Navier-Stokes equations remains relatively small when $\Re\leq 20$. By this reason constructed for $\Re\ll 1$ regular-perturbation methods~\citep{Ho:Leal74,Vass:Cox76,Cheruk:Mclau94}  have successfully predicted the lift force on a point-like neutrally buoyant particle at a moderate Re. For larger Re, when a contribution of inertial terms becomes significant, the equilibrium positions should be shifted towards the wall with the increase in $\Re$~\citep{Schon:Hinch89,Asmolov99}.

We now turn to non-neutrally buoyant particles, which density is different from that of liquid, so that they experience an external gravity force, $F'_{g}$, which in dimensionless form can be expressed as
\begin{equation}
F_{g}=\dfrac{F'_{g}}{\rho a^4 G_m^2}=\dfrac{4\pi g}{3aG_m^2}\Delta \rho, \label{eq:Fg}
\end{equation}
where $\Delta\rho=(\rho_p-\rho)/\rho$. The gravity influences both the particle migration and equilibrium position when $F_{g}=O(1)$. It also induces an additional slip velocity which is of the order of the Stokes settling velocity,
\begin{equation}
V^{St}=\dfrac{F'_g}{6\pi\mu a^2G_m}=\dfrac{\Re_p F_g}{6\pi },\label{eq:St}
\end{equation}
where $\mu $ is the dynamic viscosity. The effect of this velocity on the lift is comparable to $F_l^{nb}$ when $V^{St}=O(1)$, i.e., at large gravity, $F_g\sim 6\pi \Re_p^{-1}\gg 1$, and is very important for vertical or nearly vertical channels.
For horizontal channels, the slip velocity is equal to that of a neutrally buoyant sphere since $F_x=0$.
Equation~(\ref{our_fit}) can also be applied in this case since the slip velocity remains small far from walls. Equilibrium positions of particles, $z_{eq}$, can then be deduced from the
balance between the lift and the gravity,
\begin{equation}
c_l(z_{eq})=F_g.
\label{eq:Zeq}
\end{equation}
Equation(\ref{eq:Zeq}) may have two, one or no stable equilibrium points depending on
$F_g$, and the sensitivity of equilibrium positions to the value of $a$ or $\Delta \rho$ is defined by the value $\partial
c_l/\partial z$. Thus, when the derivative is small, small variations in $F_g$ will lead
to a significant shift in focusing positions. We finally note that the range of possible $z_{eq}$
can be tuned by the choice of $U_m'$.

%

\section{Simulation method}
\label{sec_simul}

In this section, we present our simulation method and justify the choice of parameters.

For our computer experiment, we chose a scheme based on the lattice Boltzmann method~\citep{benzi_lattice_1992,Kunert2010random,Dubov2014} which has been successfully employed earlier to simulate a motion of particles in the channel flow. We use a simulation cell confined by two impermeable
no-slip walls located at $z=0$ and $z=79\delta$, so that in all simulations $H=79 \delta$, and two
periodic boundaries with $N_{x}=N_{y}=256\delta$, where $\delta$ is the lattice
spacing. \label{add4} Spherical particles of radii $a=4\delta-12\delta$ are implemented as
moving no-slip
boundaries~\citep{ladd_lattice-boltzmann_2001,bib:jens-janoschek-toschi-2010b,HFRRWL14}, where the chosen radii are sufficient to keep discretisation effects of the order of a few percent~\citep{Janoschek2013}.
A Poiseuille flow is generated by applying a body force, which is equivalent to a pressure gradient $-\nabla p$. We use a 3D, 19 velocity, single relaxation time implementation of the lattice Boltzmann method, where the relaxation time
$\tau$ is kept to 1 throughout this paper. Different flow rates are obtained by
changing the fluid forcing.  We use two channel Reynolds numbers,
$\mathrm{Re}=11.3$ and $22.6$. To simulate the migration in an inclined
channel we apply the gravity
force directed at an angle
$\alpha$ relative to the $z$-axis at the center of the particle. In our simulations the values of dimensionless $F_{g}$ vary from $0$ (neutrally buoyant particle) to $13.91$.

In our computer experiments we determine the lift by using two different strategies. In the first method we extract the lift from the migration velocity. We measure the $x-$ and $z-$components of the particle velocity to find the dimensionless slip, $V_s=(V'_x-U'(z))/(aG_m)$, and migration velocities, $V_m=V_z'/(a G_{m})$. \label{add5} To suppress the fluctuations arising from the discretization artifacts we average the velocities over approximately 4000 timesteps. The error does not exceed $3\%$ for the particles of $a=4$ and rapidly decreases with $a$. The lift force can then be found from these calculations, by assuming that the particle motion is quasi-stationary. The
lift is balanced by the $z-$ component of the drag, $F_l'=-F_{dz}'$. Following~\citet{Dubov2014} we use an expression
\begin{equation}
F_{dz}^{\prime }\approx -6\pi \mu aV_{m}^{\prime }f_z(z/H,a/H),  \label{eq_forcefit}
\end{equation}%
\begin{equation}
f_z=1+\dfrac{a}{z-a}+\dfrac{a}{H-a-z},
\label{eq_forcefit2}
\end{equation}%
where the second and the third terms are corrections to the
Stokes drag due to hydrodynamic interactions with two channel walls. In what follows
\begin{equation}
c_{l}=6\pi V_{m}f_z\Re_p^{-1}.  \label{eq_cl}
\end{equation}
Second method to calculate the lift (and to check the validity of the
first approach) uses the balance of the lift and the gravity forces described by Equation~(\ref{eq:Zeq}). By varying the gravity force $F_g$ one can, therefore, comprehend the whole
range of equilibrium positions within the channel to obtain $c_l(z)$. The advantage of such an approach is that it does not require the particle motion to be quasi-stationary. However, the disadvantage of this method is that the convergence to
equilibrium can be slow in the central zones of the channel, where the slope of
$c_l(z)$ is small. Therefore, we use this computational strategy only  in the near-wall region.

\section{Results and discussion}

In this section, we present the lattice Boltzmann simulation results and
compare them with theoretical predictions.

\subsection{Neutrally buoyant particles}
\label{sec_nb}

\begin{figure}
 \centering
  \includegraphics[height=5.6cm]{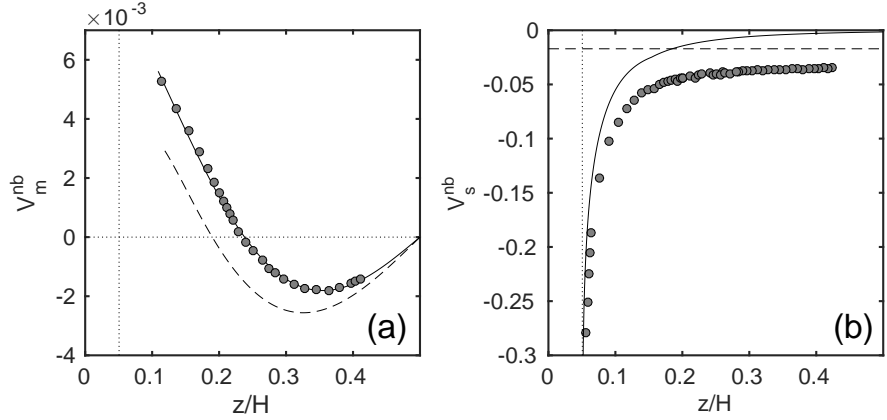}
  \caption{(a) Dimensionless  migration  velocity computed as a function of
$z/H$ for particles of $a=4\delta$ (symbols). The location of the particle in a contact with the wall, $z=a$, is shown by a vertical
dotted line. Dashed curve plots theoretical predictions for pointlike particles. Solid curve shows a polynomial fit of simulation data. (b) Dimensionless slip velocities computed for the same particles (symbols). Solid curve plots the
slip velocity in a linear shear flow near a single wall.  Dashed line plots the Faxen correction. Vertical
dotted line indicates the location of $z=a$.}
  \label{fig_vel}
\end{figure}

We start with neutrally buoyant particles and first calculate their migration $V_m^{nb}$ and the slip $V_s^{nb}$ velocities as a function of $z/H$. Figure~\ref{fig_vel} plots simulation data obtained for particles of radius $a=4\delta$. Here we show only a half of the channel since
the curves are antisymmetric with respect to the channel axis $z =H/2$. These results demonstrate that migration velocity differs significantly from the velocity $c_{l0}\Re_p/(6\pi )$, where $\Re_p=a^2G_m/\nu$, predicted theoretically for pointlike particles~\citep{Vass:Cox76}. We also see that the equilibrium position, $V_m^{nb}=0$, of finite-size particles is shifted
towards the channel axis compared to that of pointlike particles, which is obviously due to their interactions with the
wall resulting in a finite slip velocity. Indeed, Figure~\ref{fig_vel}(b) demonstrates that computed $V_s^{nb}$ grows rapidly near the
wall being close to the theoretical predictions for a linear shear flow near a
single wall~\citep{Goldman1967}. Unlike theoretical predictions by~\citet{Goldman1967}, the computed
slip velocity does not vanish in the central part of the channel. \label{add1} Its value is roughly twice larger than the Faxen correction $4U_m' a^2/(3H^2)$~\citep{happel1965}. Note that a  similar difference has been obtained in simulations of the migration of finite-size particles based on
the Force Coupling Method~\citep{Loisel2015}. These deviations from the Faxen corrections are likely also caused by hydrodynamic interactions of particles with the wall in a parabolic flow.

\begin{figure}
\centering
\includegraphics[height=5.6cm]{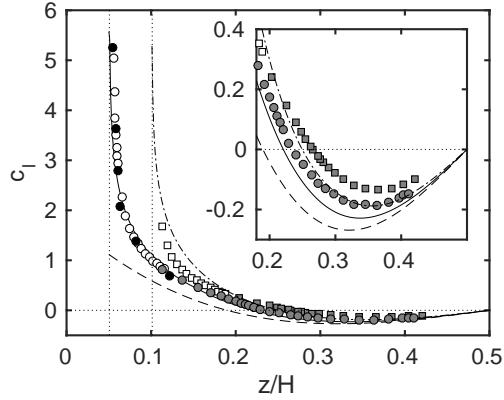}\\
\caption{Lift coefficient, $c_{l}$, for neutrally buoyant particles of $a=4\delta$ (circles) and $8\delta$ (squares) obtained from the migration velocity at $\mathrm{Re}=11.3$ (grey symbols)
and $22.6$ (white symbols).   Solid and dash-dotted curves show predictions of Equation~(\ref{our_fit})
 for $a=4\delta$ and
$8\delta$, dashed curve plots predictions for point-like particles. Vertical dotted lines show $z=a$. Black symbols show $c_{l}$ obtained for non-neutrally buoyant particles of $a=4\delta$ from the force balance at
$\mathrm{Re}=22.6$. The inset plots $c_{l}$ in the central part of the channel.
}
\label{fig_forcefit}
\end{figure}

\begin{figure}
  \centering
  \includegraphics[height=5.6cm]{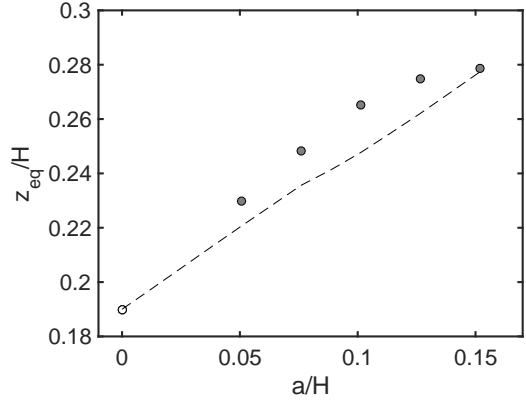}\\
  \caption{Equilibrium positions for  neutrally buoyant finite-sized
(gray circles) and point-like (white circle) particles. Dashed curve shows predictions of Equation~(\ref{our_fit}). }
  \label{fig_LH}
\end{figure}

Figure~\ref{fig_forcefit} shows $c_l$ for particles of $a=4\delta$ and $8\delta$. The lift coefficient has been obtained from the migration velocity and from the force balance as specified above, and simulations have been made for two moderate Reynolds numbers, $\mathrm{Re}=11.3$ and $22.6$. As we discussed above, if $\mathrm{Re} \leq 20$ a potential dependence of $c_l$ on $\mathrm{Re}$ could be ruled out \emph{a priori}, and this is indeed confirmed by our simulations. Therefore, below we provide a detailed comparison of our simulation data with asymptotic solutions obtained for $\mathrm{Re} \ll 1$, which should be applicable for finite moderate $\mathrm{Re}$. Figure~\ref{fig_forcefit} also includes theoretical  predictions  by \citet{Vass:Cox76} and curves calculated with Equation~(\ref{our_fit}). One can see that simulation results show strong discrepancy from point-particle approximation, especially in the near-wall region, where hydrodynamic interactions are significant. This discrepancy increases with the size of particles.  We can, however, conclude that predictions of our Equation~(\ref{our_fit}) are generally in good agreement with simulation results. Thus, for smaller particles, of $a=4\delta$, Equation~(\ref{our_fit}) perfectly fits the simulation data in the near-wall region, where the theory for point-like particles fails. Simulation results slightly deviate from predictions of Equation~(\ref{our_fit}) near the equilibrium positions and in the central part of the channel. For bigger particles, of $a=8\delta$, these deviations are more pronounced. We emphasize, however, that they are still much smaller than from the point-particle theory by \citet{Vass:Cox76}.

To examine a significance of the particle size in more detail,  we plot in Figure~\ref{fig_LH}(a) computed equilibrium position, $z_{eq}/H$, as a function of $a/H$. We recall that the lift $c_l^{nb}(z)$ is antisymmetric with respect to the midplane of the channel axis, so that neutrally buoyant particles have a second equilibrium position at $H-z_{eq}$. In a point-particle approximation $z_{eq}/H\simeq 0.19$ ~\citep{Vass:Cox76}. We see that for finite-size particles $z_{eq}/H$ is always larger, and increases with the particle size. Note that the increase in $z_{eq}/H$ is nearly linear when $a/H \leq 0.1$. Also included in Figure~\ref{fig_LH} are predictions of Equation~(\ref{our_fit}). One can conclude that the theory correctly predicts the trend observed in simulations, but slightly deviates from the simulation data. A possible explanation for this discrepancy could be effects of parabolic flow (which are of the order of $O(a/H)$)  on the slip velocity and the
stresslet~\cite[see][]{yahiaoui2010,hood2015}, which are neglected in our theory.

\subsection{Non-neutrally buoyant particles}
\label{nnb}

We now turn to the particle migration under both inertial lift  and gravity forces.

\subsubsection{Horizontal channel}

Let us start with the investigation of migration of particles in a most relevant experimentally case of a horizontal channel ($\alpha = 0^{\circ}$).

\begin{figure}
  \centering
  \includegraphics[height=5.6cm]{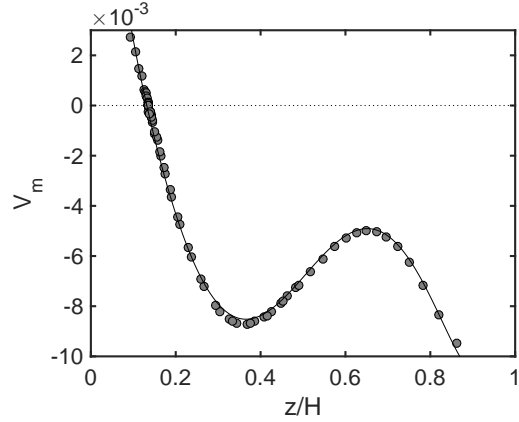}\\
  \caption{Migration velocity of non-neutrally buoyant particles in a
horizontal channel. Symbols show simulation data. Solid curve is calculation with Equation~(\ref{vm_vert}) using data for neutrally buoyant particles.}
  \label{fig_vhor1}
\end{figure}

We first fix a weak gravity force, $F_g=0.694$, and compute the migration velocity of particles of radii $a=4\delta$ in a horizontal channel. Simulation results are plotted in Figure~\ref{fig_vhor1}. We see that $V_m(z)$  is no longer antisymmetric, as it has been in the case of neutrally buoyant particles. The migration velocity can be  calculated as
\begin{equation}
V_m
=V_m^{nb}-V^{St}/f_z,
\label{vm_vert}
\end{equation}
where we use a fit for $V_m^{nb}$ computed for neutrally-buoyant particles (see Figure~\ref{fig_vel}(a)). The agreement between simulation data and calculations using Equation~(\ref{vm_vert}) is excellent, which confirms that Equation~(\ref{our_fit}) remains valid in the case of slightly non-neutrally buoyant particles. We remark that due to gravity $V_m$ is shifted downwards relative to $V_m^{nb}(z)$ shown in Figure~\ref{fig_vel}. As a result, with the taken value of $F_g$ the second equilibrium position disappeared.

We recall that this type of simulations allows one to find values of $c_l(z)$ in the vicinity of the wall by varying $F_g$. We have included these force balance results in Figure~\ref{fig_forcefit} and can conclude that they agree very well with data obtained by using another computational method and for neutrally-buoyant particles. This suggests again that above results could be used at moderate Reynolds numbers, $\Re \leq 20$, since in this case the lift coefficient does not depend on $\Re$.

\begin{figure}
  \centering
  \includegraphics[height=5.6cm]{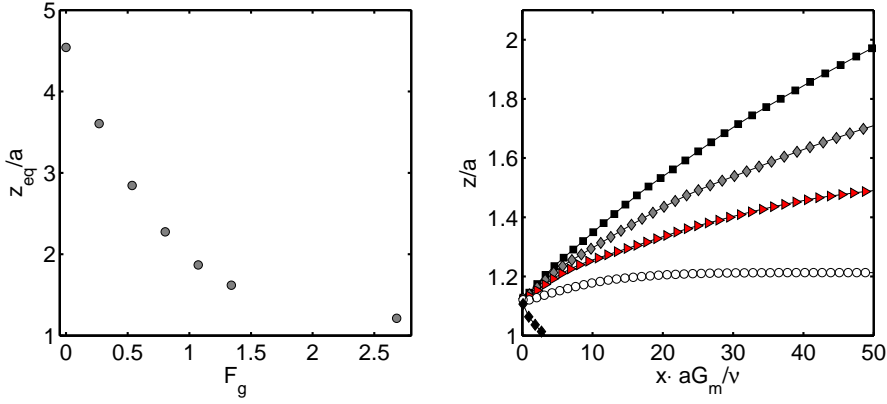}\\
  \caption{(a)~Equilibrium positions of non-neutrally buoyant particles ($a=4\delta$) in a horizontal channel; (b)~trajectories of the same particles released at $z_0=1.125 a$ computed at different $F_{g}=0.268$ (squares), $0.804$ (diamonds), $1.340$ (triangles), $2.681$ (circles), $9.383$ (diamonds). }
  \label{fig_traj}
\end{figure}

 Figure~\ref{fig_traj}(a) shows $z_{eq}/a$ computed at several $F_g$. It can be seen that when the gravity force is getting larger, the equilibrium positions decrease rapidly. This trend can be used to separate particles even when $\Delta \rho$ is very small. To illustrate this we now fix $\Re=11.3$, inject particles of $a=4 \delta$ close to the bottom of the channel, $z_0=1.125 a$, and simulate their trajectories at different $F_g$. In Figure~\ref{fig_traj}(b) we plot trajectories of particles, $z/a$, as a function of $x  G_m a \nu $. The data show that if $F_g$ is large enough, particles sediment to the wall. However, when $F_g$ is relatively small, particles follow different and divergent trajectories, by approaching their equilibrium positions. We stress that at a given $F_{g}$ and $a/H$ trajectories, shown Figure~\ref{fig_traj}(b), remain the same for any $\Re \leq 20$ (see  Appendix~\ref{app_b}). Therefore, even in the case of very small $\Delta \rho$, one can always tune the value of $\Re$ to induce the required for  separation difference in $F_g$. For example, we have to separate particles of $a=2~\mu$m and different $\Delta \rho$ in the channel of $H=40~\mu$m. If we chose $\Re=0.3$, the separation length $L=50 x  G_m a \nu$ of Figure~\ref{fig_traj}(b) will be ca. $3.3$~cm. By evaluating $\Delta \rho$ with Equation~(\ref{eq:Fg}), we can immediately conclude that trajectories plotted in Figure~\ref{fig_traj}(b) from top to bottom correspond to $\Delta \rho =0.007$, $0.022$, $0.037$ and $0.073$, which is indeed extremely small.

\subsubsection{Inclined channel}

\begin{figure}
  \centering
  \includegraphics[height=5.6cm]{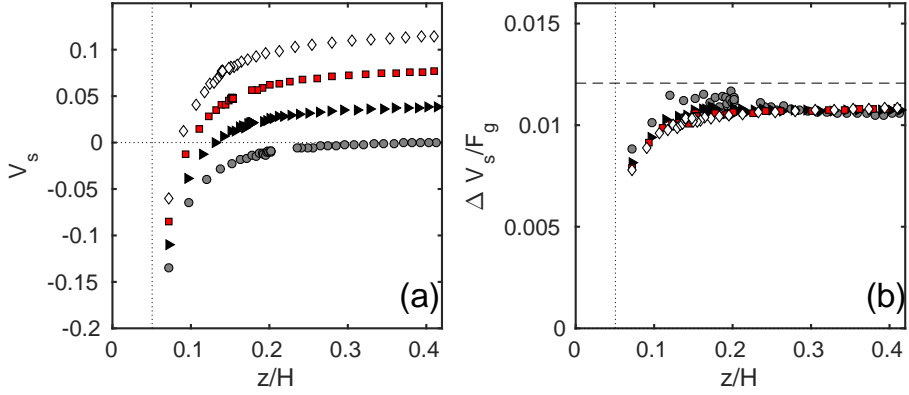}\\
  \caption{Slip velocities (a) and $\Delta V_{s}/F_g$ (b) computed for non-neutrally buoyant particles of $a=4\delta$ in a
vertical channel. The data sets correspond to $F_g=3.475$ (circles), $6.956$ (triangles), $10.44$ (squares) and $13.91$ (diamonds). Dashed line shows $V^{St}/F_g$, vertical dotted lines plot $z=a$.
}
  \label{fig_vslip_vert}
\end{figure}

\begin{figure}
  \centering
  \includegraphics[height=5.6cm]{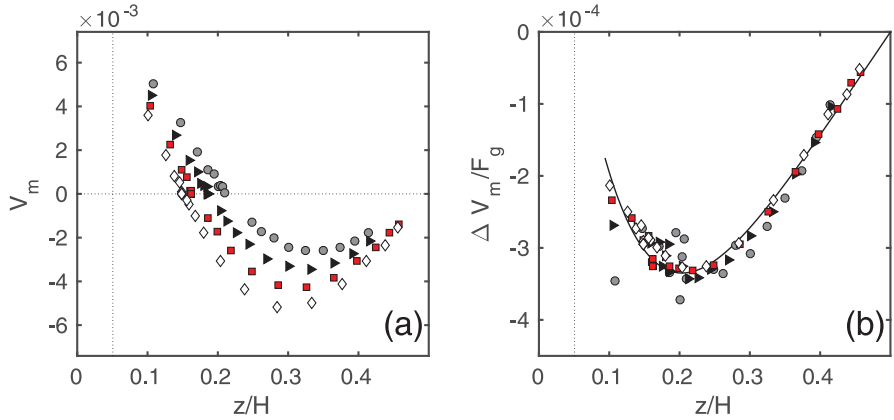}\\
  \caption{Migration velocities (a) and $\Delta V_{m}/F_g$ (b) computed for non-neutrally buoyant particles of $a=4\delta$ in a vertical channel. The data sets correspond to $F_g=3.475$ (circles), $6.956$ (triangles), $10.44$ (squares) and $13.91$ (diamonds). Vertical dotted lines plot $z=a$. Solid curve shows a polynomial fit of data.}
  \label{fig_vert}
\end{figure}

 When $F_g$ is large enough, it can also influences the slip velocity, and therefore, change the lift itself. This effect is especially important for vertical channels.
Note that due to the linearity of the Stokes equations, which govern a
disturbance flow at small particle Reynolds numbers, we can decouple the contributions of the particle-wall
interaction and of the gravity force into the slip velocity:
\begin{equation}
V_{s}=V_{s}^{nb}+\Delta V_{s}\sin\alpha,
  \label{vslip_vert}
  \end{equation}
where $\Delta V_{s}=V^{St}/f_x$ is the gravity-induced slip velocity for a vertical channel ($\alpha = 90^{\circ}$) and  $f_{x}\left( z/H,a/H\right)$ is the correction to the drag for a particle
translating parallel to the channel walls.
The  slip and the migration velocities of particles of $a=4 \delta$ in a vertical channel computed by using several values of $F_g$ are
shown in Figures~\ref{fig_vslip_vert}(a) and~\ref{fig_vert}(a). Note that the slip velocity, $V_s$, grows with $F_g$ since the Stokes velocity, $V^{St}$, is linearly proportional to $F_g$ (see Eq.(\ref{eq:St})). We now use simulation data presented in Figures~\ref{fig_vel}(a)
and~\ref{fig_vslip_vert}(a) to compute $\Delta V_{s}$, and then $\Delta V_{s}/F_g$. The results for $\Delta V_{s}/F_g$ are shown in
Figure~\ref{fig_vslip_vert}(b), and we see that all data collapse into a single curve, which confirms the validity of Equation~(\ref{vslip_vert}). Figure~\ref{fig_vslip_vert}(b) also shows that $\Delta V_{s}/F_g$ is nearly
constant in the central region of the channel, being smaller than $V^{St}$, but the deviations from $V^{St}$ grow when particles approach the wall. These results again illustrate that hydrodynamic interactions with the walls significantly affect motion of particles in the channel.

We recall that the variation of the slip velocity caused by gravity is small for slightly
non-neutrally buoyant particles (see Figure~\ref{fig_vslip_vert}), so that Eq.(\ref{our_fit}) can be linearized with respect to $\Delta V_{s}$:
\begin{equation}
c_l\simeq c_{l}^{nb}+\Delta V_{s}\frac{\partial c_{l}(V_{s}^{nb})}{\partial V_{s}},
  \label{eq_force3}
\end{equation}
where $c_{l}^{nb}=c_{l}(V_{s}^{nb})$ is the lift coefficient for neutrally
buoyant particles. By using Eq.(\ref{eq_cl}) we can then calculate the migration velocity
\begin{equation}
V_{m}=V_{m}^{nb}+\Delta V_{m}=V_{m}^{nb}+\Delta V_{s}\frac{\partial c_{l}(V_{s}^{nb})}{\partial V_{s}}\frac{\Re_p}{6\pi f_z}.  \label{eq_vm}
\end{equation}
The computed migration velocity is shown in Figure~\ref{fig_vert}(a). We see that it decreases with $F_g$, and the equilibrium position shifts towards the wall, which is since $\Delta V_{s}/F_g$ is positive while $\partial c_{l}/\partial V_{s}$ is negative.

 We can now evaluate $\Delta V_{m}/F_{g}$
by using simulation data presented in Figures~\ref{fig_vel} and~\ref{fig_vert}(a), and these
results are presented in Figure~\ref{fig_vert}(b). As one can see, the data collapse into a single curve, thus confirming the validity of our  linearization, Equation~(\ref{eq_vm}).

\begin{figure}
  \centering
  \includegraphics[height=5.6cm]{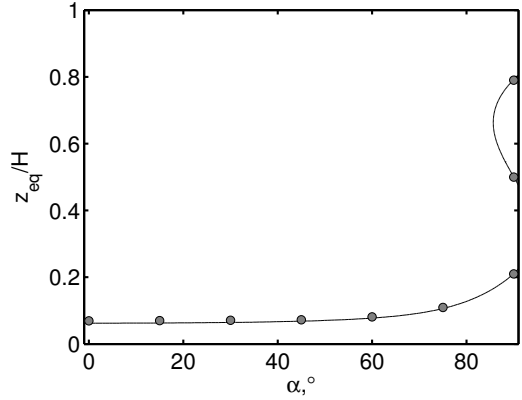}\\
  \caption{Equilibrium positions $z_{eq}/H$ for $a=4\delta$ and $F_g=3.475$. Circles show simulation data. Solid curve plots results obtained using $V_m = 0$, where $V_m$ is calculated with Eq.~(\ref{eq_vma}). }
  \label{fig_vvert2}
\end{figure}

Finally, we briefly discuss the case of an arbitrary inclination angle $\alpha$, where the $z$-component of the force can be written as
\begin{equation}
F_z= c_{l}(V_s)+F_g\cos\alpha.
\label{eq_force4}
\end{equation}
By using Eqs.(\ref{vslip_vert}), (\ref{eq_vm}) and (\ref{eq_force4}), we can express the migration velocity as
\begin{equation}
V_{m}=V_{m}^{nb}+\Delta V_{m}\sin\alpha+F_g\cos\alpha\frac{\Re_p}{6\pi f_z},  \label{eq_vma}
\end{equation}
where  $\Delta V_{m}$ is evaluated for a vertical channel (see Figure~\ref{fig_vert}(b)). The equilibrium positions can be found by using a condition $V_m = 0$, where $V_m$ is calculated with Eq.(\ref{eq_vma}). The results of these calculations  made at a  fixed $F_g=3.475$ and different $\alpha$ are plotted in Figure~\ref{fig_vvert2} together with direct simulation data, and one can see that they practically coincide. Our results show that in a vertical channel two stable equilibrium positions coexist. \label{add3} They are symmetric relative to the midplane and are located close to walls. Another, third equilibrium position has a locus at the midplane, but is unstable. A similar result has been obtained earlier~\citep{Vass:Cox76,Asmolov99}. If we slightly reduce $\alpha$ both stable equilibrium positions become shifted towards the lower wall due to gravity as well seen in  Figure~\ref{fig_vvert2}. These two positions coexist only for $\alpha\geq 85.7^{\circ}$. On reducing $\alpha$ further
the upper equilibrium position disappears, and only one, a lower, equilibrium position remains. This obviously indicates that the inertial lift cannot balance gravity anymore. We note that this remaining single equilibrium position becomes insensitive to the inclination angle when $\alpha\leq 60^\circ$.

\section{Concluding remarks}

In this paper we have studied the inertial migration of finite-size particles in a plane channel flow at moderate Reynolds numbers, $\Re\leq 20$. We have shown that the slip velocity, $V_s$, which is finite even for neutrally buoyant particles, contributes to the lift and determines the equilibrium positions in the channel. We have proposed an expression for the lift which generalizes theories, originally applied for some cases of limited guidance, to finite-size particles in a channel flow. When the size of particle turns to zero, our formula recovers known expression of a point-particle approximation~\citep{Vass:Cox76}. For particles close to the walls we recover earlier predictions for finite-size particles in a linear shear flow~\citep{Cheruk:Mclau94}.
Our theoretical model, which is probably the simplest realistic model for a lift in the channel that one might contemplate, provides considerable insight into inertial migration of finite-size particles in microchannels. In particular, it provides a simple explanation of a significant increase in the lift near walls. It also allows one to predict a number of equilibrium positions and determine their location in various situations.

To check the validity of our theory, we have employed lattice Boltzmann simulations. Generally, the simulation results have fully confirmed the theory, and have shown that many of our theoretical results have validity beyond initial restrictions of our model. Thus, it has been confirmed that predictions of our theory do not depend on Reynolds number when $\Re \leq 20$, that equilibrium positions of heavy particles in a horizontal channel can be accurately determined by using data for the neutrally buoyant case, and more.

Several of our theoretical predictions could be tested in experiment. In particular, we have shown that particles of a very small density contrast should follow divergent trajectories, so that channel flows with low Reynolds numbers $\Re\sim1$ can be used to separate such particles. \label{add7}We stress that our theory should correctly predict the lift in near-wall regions also in pipes or square channels, and we expect that for this geometry it could be accurate even at $\Re\geq 20$ since the lengthscale of the disturbance flow would rather be the distance to the wall than the channel width. By this reason it would be possible to neglect the effects of other distant walls and parabolic flow on the lift. Note, however, that these effects should be taken into account in the central part of the channel.

Our model and computational approach can be extended to more complex situations, which include, for example, hydrophobic walls or particles allowing hydrodynamic slip at their surfaces~\citep{vinogradova1999,neto.c:2005}. In this case the hydrodynamic interaction in the near-wall region changes significantly~\citep{davis1994,vinogradova1996}, so that we expect that the lift force can be also dramatically modified. It would also be interesting to consider a  case of an anisotropic superhydrophobic wall, which could induce secondary flows transverse to the direction of applied pressure gradient~\citep{feuillebois.f:2010b,vinogradova.oi:2011,harting.j:2012}. It has been recently shown~\citep{asmolov2015,pimponi.d:2014} that
particles translating in a superhydrophobic channel can be laterally displaced due to such a transverse flow. The use of this effect in combination
with the inertial migration should be a fruitful direction, which could allow to separate particles of different size or density contrast
not only by their vertical but also by transverse positions.

 ~\\
\begin{acknowledgements}
We thank Sebastian Schmieschek and Manuel Zellh\"ofer for their help on
technical aspects of the simulations. This research was partly supported by the
Russian Foundation for Basic Research (grant 15-01-03069).
\end{acknowledgements}

\appendix

\section{Fits for the slip velocity and the lift coefficients }
\label{slip}
In this Appendix we summarize known results for the slip velocity and the lift
coefficients for finite-size particles in a linear shear flow near a single
wall and for point-like particles in a Poiseuille flow. The velocity of a freely
translating and rotating particle in a linear shear flow is given
by~\citep{Goldman1967}
\begin{equation}
V_x^{nb'}=U'\left( z\right) h,
  \label{gold}
\end{equation}%
where $h$ is the correction function which depends on $z/a$ only. We use (\ref{gold}) to estimate the slip velocity in channel flow, i.e., we
neglect the effects due to parabolic flow, so that
\begin{equation}
V_s^{nb}=\frac{z(H-z)(h-1)}{aH}.
  \label{V_sh}
  \end{equation}
The correction factor fitting the results by~\citet{Goldman1967} in the
near-wall region reads \citep{reschiglian2000}
\begin{equation}
h=\frac{200.9b-\left( 115.7b+721\right) \zeta^{-1} -781.1}{%
-27.25b^{2}+398.4b-1182}\quad \mathrm{at}\quad \zeta<3,
  \label{rech}
\end{equation}%
where $\zeta = z/a$ and $b=\log (\zeta-1)$. Note that here we have reformulated the original equation~\citep{reschiglian2000} in terms of the natural logarithm.
For larger distances we use the asymptotic solution by~\citet{wakiya1967},
\begin{equation}
h=\frac{1-\frac{5}{4}\zeta^{-3}+\frac{5}{4}\zeta^{-5}-\frac{23}{48}\zeta^{-7}-%
\frac{1375}{1024}\zeta^{-8}}{1-\frac{15}{16}\zeta^{-3}+\zeta^{-5}-\frac{3}{8}%
\zeta^{-7}-\frac{4565}{4096}\zeta^{-8}}\quad \mathrm{at}\quad\zeta\geq 3.
  \label{wak}
\end{equation}

\citet{Vass:Cox76} have obtained the lift force on a particle in a channel flow
at  $\Re\ll 1$ by using a point-particle approximation:
\begin{equation}
c_{l}^{VC}=c_{l0}^{VC}+\frac{H}{a}c_{l1}^{VC}V_{s}+c_{l2}^{VC}V_{s}^{2},
\label{vc2}
\end{equation}
where coefficients $c_{l0}^{VC},c_{l1}^{VC},c_{l2}^{VC}$ depend on $z/H$ only.
Later ~\citet{feuillebois2004} has proposed a simple fitting expression:
\begin{equation}
c_{l0}^{VC}=2.25\left( z/H-0.5\right) -23.4\left( z/H-0.5\right) ^{3}.
\label{cl0}
\end{equation}
The expression for a lift coefficient of a finite-size particle in a linear shear flow near a single wall has been suggested by \citet{Cheruk:Mclau94}
\begin{equation}
c_{l}^{CM}=c_{l0}^{CM}+c_{l1}^{CM}V_{s}+c_{l2}^{CM}V_{s}^{2},
\label{cherukat}
\end{equation}%
where the coefficients $c_{l0}^{CM},c_{l1}^{CM},c_{l2}^{CM}$ depend on $\zeta$ only:
\begin{equation}
c_{l0}^{CM}=1.8081+0.879585\zeta ^{-1}-1.9009\zeta ^{-2}+0.98149\zeta ^{-3},
\label{cl0CM}
\end{equation}%
\begin{equation}
c_{l1}^{CM}=-3.24139\zeta -2.676-0.8248\zeta ^{-1}+0.4616\zeta ^{-2},
\label{cl1CM}
\end{equation}%
\begin{equation}
c_{l2}^{CM}=1.7631+0.3561\zeta ^{-1}-1.1837\zeta ^{-2}+0.845163\zeta ^{-3}.
\label{cl2CM}
\end{equation}


\section{Governing equations for trajectories of particles}
\label{app_b}
In this Appendix, we derive equations which govern particle trajectories.
The components of the particle velocity can be written as
\begin{equation}
\frac{dx}{dt}=V_{x}^{\prime }=U^{\prime }\left( z\right) h=G_mz\left(
1-z/H\right) h,  \label{3}
\end{equation}%
\begin{equation}
\frac{dz}{dt}=V_{m}^{\prime }=\frac{F_{l}^{\prime }-F_{g}^{\prime }}{6\pi
\mu af_z}=\frac{\left( c_{l}-F_{g}\right) a^{3}G_m^{2}}{6\pi \nu f_z  }.  \label{4}
\end{equation}
The last equality indicates that the migration time, i.e., the time required for a particle to
migrate at distance of the order of its radius $a$, is equal to $\nu /\left( G_ma\right)
^{2}=(4G_m\Re)^{-1}(H/a)^2.$ Since the right-hand sides of~(\ref{3}) and~(\ref{4}) do not explicitly include
time, one can formulate an equation governing the particle trajectory
as
\begin{equation}
\frac{dz}{dx}=\frac{a^{3}G_m}{6\pi \nu }\frac{c_{l}-F_{g}}{f_zz\left(
1-z/H\right) h}.  \label{zx}
\end{equation}%

Let us now turn to dimensionless coordinates $\zeta$ and $\xi =xG_ma/\nu$. Equation~(\ref{zx})
can then be rewritten as
\begin{equation}
\frac{d\zeta }{d\xi }=\frac{1}{6\pi }\frac{c_{l}-F_{g}}{f_zh\zeta \left(
1-\zeta a/H\right) }.  \label{6}
\end{equation}%

We stress that Equation~(\ref{6}) does not depend on $\Re$. Indeed,  $f_z$ and $h$ are dimensionless functions of $\zeta $ and $a/H$ only, and the lift coefficient, $c_{l}$, is also not sensitive to the Reynolds number when $\Re \leq 20$. This implies that at given $F_{g}$ and $a/H$ trajectories satisfying Equation~(\ref{6}) are universal, i.e. remain the same at any $\Re \leq 20$.




%
\bibliographystyle{jfm}
\bibliography{lift}

\end{document}